 \documentclass[%
 reprint, aps]{revtex4-1}

\usepackage{booktabs}
\usepackage{xcolor}
\usepackage{graphicx}
\usepackage{graphics}
\usepackage{amsmath}
\usepackage{amssymb}
\usepackage{amsthm,amstext}
\usepackage{amsfonts}
\usepackage{lipsum}
\usepackage{MnSymbol}
\usepackage{mathrsfs}
\usepackage{stmaryrd}
\usepackage{latexsym}
\usepackage[applemac]{inputenc}
\usepackage{accents}
\newcommand\thickbar[1]{\accentset{\rule{.4em}{.8pt}}{#1}}

\usepackage[bbgreekl]{mathbbol}
\DeclareSymbolFontAlphabet{\mathbbm}{bbold}
\DeclareSymbolFontAlphabet{\mathbb}{AMSb}%

\newcommand\bb{\textbf{\emph{b}}}

\newcommand\bv{\textbf{\emph{v}}}
\newcommand\bu{\textbf{\emph{u}}}

\newcommand\bx{\textbf{\emph{x}}}

\newcommand\bs{\textbf{\emph{s}}}
\newcommand\bn{\textbf{\emph{n}}}
\newcommand\bw{\textbf{\emph{w}}}

\newcommand\0{\textbf{\emph{0}}}

\newcommand\I{\textbf{I}}

\renewcommand\d\delta
\newcommand\D\Delta
\newcommand\e{\varepsilon}
\newcommand\s{\sigma}

\newcommand\btau{\boldsymbol{\tau}}

\newcommand\bsigma{\boldsymbol{\sigma}}

\newcommand\bet{\boldsymbol{\eta}}

\newcommand\beps{\boldsymbol{\epsilon}}

\newcommand\sym{\text{sym}}

\renewcommand\div{\text{div}}

\newcommand\curl{\text{curl}}

\newcommand\Body{\mathscr{B}}

\newcommand\scrU{\mathscr{U}}

\newcommand\bbP{\mathbb{P}}
\newcommand\bbC{\mathbb{C}}

\newcommand\beq{\begin{equation}}
\newcommand\beqn{\begin{eqnarray}}
\newcommand\eeq{\end{equation}}
\newcommand\eeqn{\end{eqnarray}}

\newcommand{\trsp}{^{\hspace{-1pt}\textsf{T}\hspace{-1pt}}}

\usepackage[normalem]{ulem}

\begin{document}

\title{Inelastic Surface Growth}

\author{Giuseppe Zurlo}%
\email{giuseppe.zurlo@nuigalway.ie}
\affiliation{ School of Mathematics, NUI Galway, University Road, Galway, Ireland.}%
\author{Lev Truskinovsky}
\email{trusk@lms.polytechnique.fr}
\affiliation{PMMH, CNRS UMR 7636, PSL, ESPCI,
10 rue de Vauquelin, 75231 Paris, France.}


\begin{abstract}	

Inelastic surface growth associated with continuous creation of incompatibility on the boundary of an evolving body is behind a variety of natural and technological processes, including embryonic development and  3D printing. In this paper we extend the recently proposed stress-space-centered theory of  surface growth (PRL 119, 048001, 2017) by shifting the focus towards growth induced strains. To illustrate the new development we present several analytically tractable examples.

\begin{center}
\emph{The paper is dedicated to the memory of G{\'e}rard Maugin, a scientific visionary. }
\end{center}

\end{abstract}

\maketitle

\section{Introduction}

In the process of   surface  growth  the configuration of a body is changing as a result of continuous deposition  of new matter on its boundary \cite{Skalak, DiCarlo, MauginCiarletta, Goriely17,Tomassetti}. Despite the ubiquity of such processes  in both living and inert systems,  the mechanical theory of surface growth, accounting for the development of inelastic strains in the growing body,  is still far from being complete. This is rather remarkable  given that the mechanical theory of   volumetric growth, assuming that  mass supply takes place  in the bulk of the body,  is  well advanced  \cite{MauginEpstein, MauginCFG, Yavari, Kuhl,  BenAmar,  Ateshian, Ambrosi, ChapmanJones}. 

 The approach to  inelastic surface growth developed   in the present paper is focused on the elastic incompatibility created at the instant of deposition. The associated inelastic strain is not  evolving after  accretion and therefore  depends exclusively on the deposition protocol. Recently, relying on some previous insights \cite{Goodman, Fletcher, Trincher, Naumov, Arutyunyan, Gambarotta, SozioYavari, Papadopoulos}, we were able to link explicitly the  deposition strategy with the final incompatibility \cite{ZT17}. However, the proposed description was limited to the stress space, which was possible because the growing surface was assumed to be loaded in a soft device.
 
To handle the  case of a hard device,  one needs to  address  the kinematical issues that were  bypassed in the stress-centered formulation. This is the goal of the present paper where we introduce  the instantaneous displacement field  for the particles materializing on the growing surface, allowing us to separate   elastic and  inelastic strains.  We distinguish further between  compatible and incompatible inelastic strains and we link the latter to the accumulation of residual stresses.

As an illustration, we consider a bar growing by one of its ends. To see the  development of  incompatibility   in this minimal setting we  laterally constrain  the bar  by attaching it to an elastic background. We show that even in this elementary case  the obtained solutions depend nontrivially on  deposition protocols, suggesting interesting applications  for reinforced masonry and   additive manufacturing.

Throughout the paper we assume  that  strains are small and  that dynamical effects can be neglected. A more general theory will be presented elsewhere.

\section{General theory\label{SecGen}}
 
\emph{Kinematics.} 
In order to track the evolution of material points after accretion, we introduce a  reference  mass reservoir  $\Body_t=\left\{\bx\,|\,\vartheta(\bx)\leq t\right\}$.  For simplicity, in this study the scalar function $\vartheta$ will be prescribed, even though in many realistic growth processes its evolution would have to be found self-consistently. The Lagrangian velocity of $\partial\Body_t$  is then  $D_0=(\nabla\vartheta(\bx)\cdot\bn_0(\bx))^{-1}$   where $\bn_0=\nabla\vartheta(\bx)/|\nabla\vartheta(\bx)|$ is the reference outward normal. We can decompose  $\partial\Body_t=\Omega_t\cup\Omega_d$, where the time dependent growing part $\Omega_t$ is such that $D_0\neq 0$ while the time independent non-growing part $\Omega_d$ has $D_0=0$. Writing the incremental advance of the material surface $\Omega_t$ as   $\D\bx=D_0\bn_0\D t$, where $\D t$ is an infinitesimal time interval,  we obtain $j_0=\rho_0 D_0$, where $j_0$ is the mass flux per reference surface and  $\rho_0(\bx)$ is the time-independent volumetric mass density in $\Body_t$.  Since the reference configuration   is defined only for the material points already in  $\Body_t$, it can be always chosen to coincide with the  instantaneous actual placement of the body  in $\textbf{R}^3$. 

For each instant $t\geq\vartheta(\bx)$, we can introduce displacement of the attached material particles 
 $\bu(\bx,t)$. In particular,  the instantaneous (initial)  displacement  at the moment of deposition is 
$
\mathring\bu(\bx)=\bu(\bx,\vartheta(\bx)).
$
The presence of this displacement field makes the Eulerian normal velocity $D$ of the accreting surface in the actual space $\omega_t $ different from the Lagrangian velocity $D_0$.  More specifically,  we can write  $D =D_0+D_0(\nabla\mathring\bu)\bn_0\cdot\bn$, where $\bn$ is the normal to $\omega_t$. 
 
The difference  between $D$ and $D_0$ plays a fundamental role in surface growth. For instance, in the process of 3D printing, the velocity of the printing head  can be controlled independently of the  mass flux. Since in this paper  we neglect geometrical nonlinearities, we can assume that $\bn_0\sim\bn$ and write the relation between $D_0$ and $D$ in the form $D/D_0-1=(\nabla\mathring\bu)\bn\cdot\bn. $

\bigskip
 
\emph{Elastic growth.} Assume that the reference configuration $\Body_t$ is a stress free state for the growing body. If body forces $\bb$ and surface tractions $\bs$ are controlled during growth, equilibrium of intermediate configurations requires that, for all  $ t\in(0,T)$,  where $T$ corresponds to the end of accretion, 
\beq\label{equi}
\left\{
\begin{array}{lcc}
\div\bsigma(\bx,t) + \bb(\bx) =\0 & \text{in} & \Body_t\\
\bsigma(\bx,t)=\bbC(\bx)\nabla_{\hspace{-1pt}s}\bu(\bx,t) &\text{in} & \Body_t\\
\bsigma(\bx,t)\bn(\bx) =\bs(\bx)  & \text{on} & \Omega_t\\
\bu(\bx,t)=\0 & \text{on} & \Omega_d
\end{array}
\right.
\eeq
where  $\bsigma(\bx,t) $ is the (symmetric) Cauchy stress tensor, $\bbC$ is a positive definite and symmetric elasticity tensor and  $\nabla_{\hspace{-1pt}s}=(\nabla+\nabla\trsp)/2$. Here we have assumed that the material behavior of the body is linearly elastic and that the strains are fully defined by the gradient of the displacement field $\bu(\bx,t)$.  

For each $t\in(0,T)$ the problem \eqref{equi} admits a unique solution. In anticipation of what follows,  we note that one can use \eqref{equi} to obtain the ``deposition protocol'' represented, for instance,  by the combination of  boundary displacements and surface stresses on $\Omega_t$, 
\beq\label{su}
\left\{
\begin{array}{lcc}
\mathring{\bu}(\bx)=\bu(\bx,\vartheta(\bx))\\
\mathring{\bsigma}_a(\bx)=\bbP(\bx)\bsigma(\bx,\vartheta(\bx))\bbP(\bx)
\end{array}
\right.
\quad\bx\in\Omega_t
\eeq
where $\bbP=\I-\bn\otimes\bn$ is the surface projector. From  \eqref{su}  one can also compute  the current velocity $D$, which in general will  be different from $D_0$.  
\bigskip

\emph{Inelastic growth.} To allow for inelastic growth we modify the equilibrium problem as follows
\beq\label{equiplast}
\left\{
\begin{array}{lcc}
\div\bsigma(\bx,t) + \bb(\bx) =\0 & \text{in} & \Body_t\\
\bsigma(\bx,t)=\bbC(\bx)\left(\nabla_{\hspace{-1pt}s}\bu(\bx,t)-\beps_p(\bx)\right) &\text{in} & \Body_t\\
\bsigma(\bx,t)\bn(\bx) =\bs(\bx)  & \text{on} & \Omega_t\\
\bu(\bx,t)=\0 & \text{on} & \Omega_d. 
\end{array}
\right.
\eeq
Here we still assume that the elastic response of the body is Hookean, but we introduce the inelastic strain $\beps_p(\bx)$. As it was first observed in a biological setting \cite{Nikitin}, the inelastic strain defines a relaxed (stress-free) configuration which may not be isometrically embeddable into $\textbf{R}^3$.  

The  linear elasticity problem \eqref{equiplast} is under-determined since we still did not prescribe the procedure to  find the six unknown functions  $\beps_p(\bx)$. The underlying continuous  degeneracy of the elastic energy is due to the assumption that  inelastic strains are not controlled self-consistently in the bulk of the body by, say, a flow rule of some kind, but instead these are prescribed rigidly at the moment of deposition. 
To  deal with this degeneracy, we need to specify the deposition protocol and here it will be convenient to distinguish between a ``direct problem'', where $\beps_p$ is controlled by the appropriately chosen   conditions on the accretion boundary, and an  ``inverse problem'',  where a target $\thickbar\beps_p$ is prescribed while  the associated deposition protocol is to be found. 

\bigskip

\paragraph{} If the field $\thickbar{\beps}_p(\bx)$ is prescribed the problem \eqref{equiplast}, or its modification with displacements boundary conditions on the growing surface, can be solved for each $t\in(0,T)$.   Then, by evaluating the resulting displacement and stress on $\Omega_t$ we can find the ``deposition protocol'', exemplified by the couple of functions   $ \mathring{\bu}(\bx)$ and $\mathring{\bsigma}_a(\bx)$.  In particular, for the elastic growth with $\thickbar{\beps}_p={\bf 0}$, the prescription of  $\vartheta(\bx)$, $\bb(\bx)$ and $\bs(\bx)$ gives a unique distribution  $ (\mathring\bu ,\mathring{\bsigma}_a)$.   

\bigskip

\paragraph{} If the functions $\beps_p(\bx)$ are unknown, 6 supplementary conditions are required to close the problem. Such conditions may take different forms depending on the specific aspects of mass deposition strategy. For instance,  we can prescribe the  {\it mixed} supplementary conditions in the form 
\beq\label{suplast}
\left\{
\begin{array}{lcc}
\bu(\bx,\vartheta(\bx)) = \mathring{\bu}(\bx)\\
\bbP(\bx)\bsigma(\bx,\vartheta(\bx))\bbP(\bx) = \mathring\bsigma_a(\bx) 
\end{array}
\right.
\quad\bx\in\Omega_t
\eeq
where now $\mathring\bu(\bx)$  is a  prescribed displacement of the accreting surface, and  $\mathring\bsigma_a$ is a  prescribed  active surface stress, assumed to be symmetric and satisfying $\mathring\bsigma_a\bn=\0$. 

Note that the term  ``active'' is used to distinguish $\mathring\bsigma_a$  from the conventional ``passive''   stress $\mathring\bsigma_s=\bs\otimes\bn+\bn\otimes\bs-(\bs\cdot\bn)\bn\otimes\bn$. The whole  stress tensor $\mathring\bsigma(\bx)=\bsigma(\bx,\vartheta(\bx))=\mathring\bsigma_a(\bx)+\mathring\bsigma_s(\bx)$ is therefore controlled on  the accreting boundary which makes the corresponding  elasticity problem unusual (cf. \cite{Trincher, Gambarotta}). It is even more unconventional since we can also prescribe the displacement field $\mathring{\bu}(\bx)$. Behind this ``freedom''  is, of course,  the extreme degeneracy of the elastic energy allowing for unlimited ``fluidity''  of  the arriving material.

\bigskip

\emph{Non-incremental approach. } As we have already mentioned, with  $\beps_p(\bx)$ given,  the  linear elastic problem defined by \eqref{equiplast}-\eqref{suplast} can be solved uniquely.  The solution can be written in the form $\bu(\bx,t)=\scrU(\bx,t; \beps_p)$, where we have indicated the parametric dependence of the solution on $\beps_p(\bx)$.

If $\beps_p(\bx)$ in unspecified, to find this field we need to use the solution of the linear problem $\bu(\bx,t)=\scrU(\bx,t; \beps_p)$ together with \eqref{suplast}, which leads to a time-independent system of 6 partial differential equations. It can be solved if the appropriate boundary conditions are provided, giving for instance the values of $\beps_p$ on the initial domain. 

Note that our ``mixed'' protocol  \eqref{suplast} is not the only possibility, for instance, instead of  $\mathring\bu$,  we may   choose to control only the normal velocity $D$, which gives 1 scalar condition on the normal component of the derivative of $\mathring\bu$. The remaining 2 conditions would have to reflect the microscopic details of the deposition process.

\bigskip

\emph{Incremental approach. } Following \cite{ZT17} we now consider an alternative  incremental  formulation of the same problem which does not  refer to the plastic strain $\beps_p(\bx)$ explicitly.   We begin by  introducing incremental stress  field $\dot\bsigma(\bx,t)$. Then the
 the total stress at time $t$ is
\beq\label{totstress}
\bsigma(\bx,t) = \mathring\bsigma(\bx) + \int_{\vartheta(\bx)}^t\dot\bsigma(\bx,z)\,dz.
\eeq
Field equations for the incremental displacements are found by time differentiation of \eqref{equiplast}$_{1,2,4}$,  which leads to  a sequence of incremental  problems  for  $t\in(0,T)$
\beq\label{dotred}
\left\{
\begin{array}{lrrr}
\div\dot\bsigma(\bx,t)=0 & \Body_t\\
\dot\bsigma(\bx,t)=\bbC\nabla_{\hspace{-1pt}s}\dot\bu(\bx,t) & \Body_t\\
\dot\bu(\bx,t) = \0 & \Omega_d,\\
\end{array}
\right.
\eeq
where $\dot\bu(\bx,t)$ is  incremental  displacement  field. To close  each of these equilibrium systems we need a condition on the growing part of the boundary.  Since here $\bsigma(\bx,\vartheta(\bx))\equiv\mathring\bsigma(\bx)$, we can take  a divergence of this relation to obtain the Hadamard  relation (see \cite{ZT17} for more detail)
\beq\label{HadaGen}
\div\bsigma\big{|}_{\Omega_t} = \div\mathring\bsigma - D_0^{-1}\dot\bsigma\big{|}_{\Omega_t}\bn. 
\eeq
Combined  with \eqref{equiplast}$_1$, this identity  delivers the   
condition on on $\Omega_t$  prescribing the incremental tractions  \cite{Trincher}
\beq\label{Trincher}
\dot\bsigma(\bx,\vartheta(\bx))\bn(\bx) = D_0(\bx)\left[\div\mathring\bsigma(\bx) + \bb(\bx)\right].
\eeq
Observe the appearance of the body force 
 $ \div\mathring\bsigma+\bb$, 
which represents a  fictitious pre-stressing of the incoming material. 

By solving this set of incremental problems  we can find the stress $\bsigma(\bx,t)$ but we still need to find  the inelastic strain $\beps_p(\bx)$.  However, since we  know  
the  incremental  displacement  field  $\dot\bu(\bx,t)$ and the instantaneous displacement field at the moment of deposition $\mathring\bu(\bx)$, we can   compute the total  displacement  field
\beq\label{uuu}
\bu(\bx,t) =  \mathring\bu(\bx) + \int_{\vartheta(\bx)}^t\dot\bu(\bx,z)\,dz.
\eeq
Then, by differentiating the identity  $\bu(\bx,\vartheta(\bx))=\mathring\bu(\bx)$   we   obtain 
\beq
\nabla\bu\big{|}_{\Omega_t} = \nabla\mathring\bu -\dot\bu\big{|}_{\Omega_t}\otimes\nabla\vartheta,
\eeq
which, combined   with \eqref{equiplast}$_2$,  finally gives 
\beq\label{eplastic}
\beps_p(\bx)= \sym\left( \nabla\mathring\bu(\bx)- \dot\bu\big{|}_{\Omega_t}\otimes\nabla\vartheta(\bx)\right) - \bbC^{-1}\mathring\bsigma(\bx). 
\eeq
To compute   $\dot\bu\big{|}_{\Omega_t}=\dot u(\bx,\vartheta(\bx))$ in  \eqref{eplastic} we need to know   the incremental solution at $t=\vartheta(\bx)$ which makes this ``constitutive'' relation  history dependent.

\bigskip

 \emph{Incompatibility}. The accumulated plastic strains affect both the distribution of residual stresses and the final shape of the body. To elucidate this aspect, we introduce the {\it incompatibility} tensor
\beq\label{constraint0}  
 \bet(\bx) = - \curl\curl\beps_p(\bx), 
\eeq
whose role  in the context of surface growth was discussed in detail in \cite{ZT17}, see also \cite{Naumov,Gambarotta}. For instance, this tensor  (which is symmetric and satisfies $\div\bet=\bf 0$) is the only source of residual stresses $\btau$ in a unloaded body $\Body$, which can be determined by solving the system
\beq\label{res}
\left\{
\begin{array}{llll}
\div\btau(\bx) =\0 & \text{in}  & \Body\\
\curl\curl\bbC^{-1}(\bx)\btau(\bx) = \bet(\bx) & \text{in}  &  \Body\\
\btau(\bx)\bn(\bx) =\0  & \text{on} & \partial\Body. 
\end{array}
\right.
\eeq
If at the end of accretion $\bet=\bf 0$, the  distribution of plastic strain is compatible, in the sense that  there exists a vector field $\bv$ such that
$
\beps_p(\bx)=\nabla_s\bv(\bx). 
$

To illustrate the role of the compatible component of plastic strain $\nabla_s\bv$, consider a stress-free body $\Body$ which  is fixed on   $\Omega_d$, while, for simplicity,  being traction free on $\Omega_t$.  By introducing the displacement field  $\bw=\bu-\bv$, where  $\bu$ solves  \eqref{equiplast},  we obtain
\beq\label{compeq}
\left\{
\begin{array}{lcccc}
\div\bsigma(\bx) =\0 & \text{in} & \Body\\
\bsigma(\bx)=\bbC(\bx)\nabla_s\bw(\bx) & \text{in} & \Body\\
\bw(\bx) = -\bv(\bx) & \text{on} & \Omega_d\\
\bsigma(\bx)\bn(\bx) =\0  & \text{on} & \Omega_t.
\end{array}
\right.
\eeq
If $\bv=\0$ on $\Omega_d$  we obtaine  $\bw=\0$ which means that the body is stress-free. If instead $\bv\neq\0$ on $\Omega_d$, the body will be stressed and, in particular, there will be reactive forces exerted by this surface. If we  detach the body from the constraint,  it  will change its shape as  these stresses will relax. Linking  such compatible inelastic strains with potentially complex relaxed shapes  presents an interesting challenge in applications \cite{Danescu}. 
 
 \bigskip
 
\emph{Compatible and incompatible growth.}  We define ``compatible''  growth by the condition that  the final configuration has $\bet\equiv\bf 0$. Otherwise, the  growth will   be ``incompatible'' and the outcome depends \emph{a priori} on both controls  $\mathring\bu(\bx)$ and $\mathring\bsigma_a(\bx)$. However, as it was shown in \cite{ZT17},  only the latter affects the field $\bet(\bx)$. Indeed, if we use  \eqref{eplastic}, we obtain 
\begin{equation}\label{main}
\bet = \mathring\bet-\nabla\vartheta\times\left[\curl(\bbC^{-1}\dot\bsigma)\right]_{\Omega_t}\trsp - \curl\left[\nabla\vartheta\times(\bbC^{-1}\dot\bsigma\big{|}_{\Omega_t})\right]
\end{equation}
where   $\mathring\bet  = \curl\,\curl(\bbC^{-1}\mathring\bsigma)$ is the ``arriving''  incompatibility. The right hand side of \eqref{main} does not depend on    $\mathring\bu(\bx)$, which only affects  the compatible part of inelastic strain represented by the field $\bv(\bx)$.

The fact that incompatibility is independent of $\mathring\bu(\bx)$ is important when the controlled surface growth targets a particular distribution  $\thickbar\bet(\bx)$. In this case, the problem can be fully confined in the stress space \cite{ZT17}. For instance, if the whole surface of the growing body is unconstrained $(\Omega_d=\{\O\})$,  the problem \eqref{equiplast}-\eqref{suplast} can be formulated in a displacement-free form 
\beq\label{purestress}
\left\{
\begin{array}{lcrr}
\div\bsigma(\bx,t) + \bb(\bx)=0 & & \Body_t\\
\curl\curl\bbC^{-1}\bsigma(\bx,t)=\thickbar\bet(\bx) & & \Body_t\\
\bsigma(\bx,t)\bn_0(\bx)= \bs(\bx) & &  \Omega_t.
\end{array}
\right.
\eeq
The solution to such generalized Beltrami-Mitchell problem \cite{Gupta} gives the stress field $\bsigma(\bx,t)$, whose trace on the accreting surface defines corresponding protocol $\mathring\bsigma_a(\bx) = \bbP(\bx)\bsigma(\bx,\vartheta(\bx))\bbP(\bx)$.

\section{ 1D problem}

As a first example,  consider the growth of a  bar  at one of its ends. While this case is oversimplified  because  any 1D distribution of plastic strain  is   integrable,   the  accumulation of compatible  inelastic strain can still take place during surface deposition. 

We define the reference domain of the bar as $\Body_t=\left\{x\in(0,\psi(t))\right\}$, where the prescribed monotone function $\psi(t)$ defines the position of  the growing end.  The  Lagrangian  velocity of the accreting front at $x$ is  then $D_0(x)=1/\vartheta'(x)$, where $\vartheta=\psi^{-1}$. The Eulerian velocity is 
$D(x)=D_0(x)(1+{\mathring u}'(x))$ where   $\mathring u(x)=u(x,\vartheta(x))$ is the scalar analogue of  \eqref{su}$_1$. Note that in this case $\mathring{u}$ is completely defined by $D/D_0$ through
\beq\label{uDD}
\mathring{u}(x) = \int_0^x\left(D(z)/D_0(z) - 1 \right)\,dz
\eeq
where we have set $\mathring{u}(0)=0$. 

Denote by $\s(x,t)$ the stress in the growing bar.  If the end $x=0$ is fixed, the equilibrium conditions under body forces $b(x)$ and  tractions  on the growing-end $s(\psi(t))$   take the form 
\beq\label{equiplast1D}
\left\{
\begin{array}{llll}
\partial_x\s(x,t) + b(x) = 0 & \Body_t\\
\s(x,t) = E(\partial_x u(x,t) - \e_p(x)) & \Body_t\\
\s(\psi(t),t) = s(\psi(t)) & \Omega_t\\
u(0,t) = 0 & \Omega_d. 
\end{array}
\right.
\eeq
This system is the 1D analog of \eqref{equiplast}. If the   time-independent plastic strain  $\e_p(x)$ is viewed as prescribed, the problem \eqref{equiplast1D} can be solved explicitly 
\beqn
&& u(x,t) =  x\left(\frac{s(\psi(t))}{E}-\e_p(\psi(t))\right) \nonumber\\
&& \hspace{30pt}+ \int_0^x\hspace{-7pt}\int_v^{\psi(t)}\left(\frac{b(z)}{E}+\e_p'(z)\right)dz dv. 
\eeqn
From this expression we can  compute the field $\mathring u(x)$  which is determined by the tractions, controlled on $\Omega_t$. We can also obtain the expression for the Eulerian velocity of the growing end 
\beq\label{1D}
\frac{D}{D_0} - 1 = {\mathring{u}}' = \e_p + \frac{s}{E} + \frac{x}{E}\left(s'+b\right). 
\eeq
 which is controlled by the  three factors:   the imposed plastic deformation,  the traction-induced elastic pre-strain in the incoming material  and the body forces acting on the arriving material points. It is clear that instead of prescribing surface traction $s(x)$  and finding  surface velocity  $D(x)$, we could \emph{vice versa} prescribe the velocity and find the traction.
 
 If now  the function $\e_p(x)$ is viewed as unknown, the problem  must be completed with a single supplementary condition. Since there is no analogue in 1D of the surface stress $\mathring\bsigma_a$, the only option is to specialize $\mathring{u}(x)$ or, equivalently,  ${\mathring u}'=D/D_0-1$, from which we find the plastic strain  $\e_p = {\mathring{u}}' -  s/E -  x\left(s'+b\right)/ E$. In particular, if $s=b=0$, we obtain that $\e_p(x)\equiv{\mathring{u}}'(x)$, which shows that  $\mathring{u}$  is the    displacement field connecting the two stress free configurations of the bar: the initial and the final ones.  
 
\begin{figure}[!th]
\includegraphics[scale=.35]{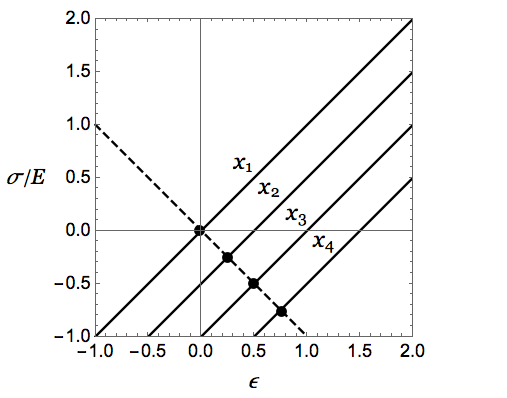}
\caption{\label{baptism} The constitutive behavior of the deposited material. The dashed line represents the ``yield curve'', relating the initial stress with the total strain at deposition. The solid lines represent the elastic  stress-strain curves for  4 material points deposited at $x_1=0$, $x_2=0.25$, $x_3=0.5$, $x_4=0.75$. They keep the memory of the initial state at deposition  shown by the black circles. Here $L=1$, $C=-1$.}
\end{figure}

We can use this example to show that, in contrast to classical plasticity,  our ``pseudo-plastic'' material behaves elastically while being both loaded and unloaded.  Assume that during growth we impose, at the same time, both the velocity $D=D_0$ and the space dependent tractions $s(x)=C x$, with $C$ a constant. In the absence of body forces we thus obtain $\e_p(x)=-2 C x / E$. Then for $t\geq\vartheta(x)$  the constitutive response is space dependent  $\s(x,t) = E(\e(x,t) + 2 C x/E)$, where $\e=\partial_x u$ is the total strain. We illustrate this  stress-strain behavior in
Fig.\ref{baptism}, where the stress-strain relation at deposition, denoted by a dashed line, is the analog of  the ``yield curve'', since after deposition the material elements behave elastically.  

\section{ 1.5D problem }

The goal of  our second example is to   create room for incompatibility without compromising the analytical transparency of the 1D setting. To this end we assume that the bar is laterally constrained by an elastic background, composed  of leaf (shear) springs attached to a rigid wall.  We will refer to such compound system as  a ``1.5D  bar'' because the problem remains to  be  ODE based. 

\bigskip

\emph{Incompatibility.}  Consider  first a fixed length 1.5D bar  in a stress-free reference configuration $\Body=\left\{x\,|\,0\leq x\leq L\right\}$. Denote by  $u(x)$ the displacement field  in $\Body$ and by $\mathring{u}$ the shift  of the attachment points of the shear springs relative to the background. Then the bar is subjected to a distributed live loading by the body forces
$
q(x)=-k(u(x)-\mathring{u}(x)). 
$
If no other loads act on the bar,  equilibrium equations read 
\beq\label{15Dfirst}
\left\{
\begin{array}{lll}
\s'(x) + q(x) = 0 \\
\s(0)=\s(L)=0. 
\end{array}
\right.
\eeq
Supplementing this system by the constitutive relations 
$
\sigma(x)= E (u'(x)-\e_p(x)) 
$
and eliminating  $u$  we obtain 
\beq\label{15Dsecond}
\s''(x) - \beta^2\s(x)  = k(\e_p(x) - \mathring{u}'(x))
\eeq
where we have set $\beta^2=k/E$.  Observe that  $ \beta$ has a dimension  of   inverse length scale and   that  the dimensionless ratio  $\bar\beta = \beta L$  characterizes  the role of  elastic foundation: if  $\bar\beta$ is small, the  foundation can be neglected and if   $\bar\beta$ is large the foundation dominates the response.  

Observe also that Eq.  \eqref{15Dsecond} is the 1.5D counterpart of \eqref{res}$_{1,2}$, and that the function 
\beq\label{inc15}
\eta(x)=\e_p(x)-\mathring{u}'(x)
\eeq
can be viewed as the analog of  incompatibility in this problem:  a measure of the  mismatch between  the ``external" (non-1D)  and the ``internal"  (1D)  pre-strains, cf.   \cite{Basile17}. For instance, the residual stresses in this setting are defined by $\eta(x)$ rather than $\e_p(x)$ and even if  $\e_p=0$  there may be equally ``inelastic"  strains  $\mathring{u}'$ in the system.

\begin{figure}[!th]
\includegraphics[scale=.23]{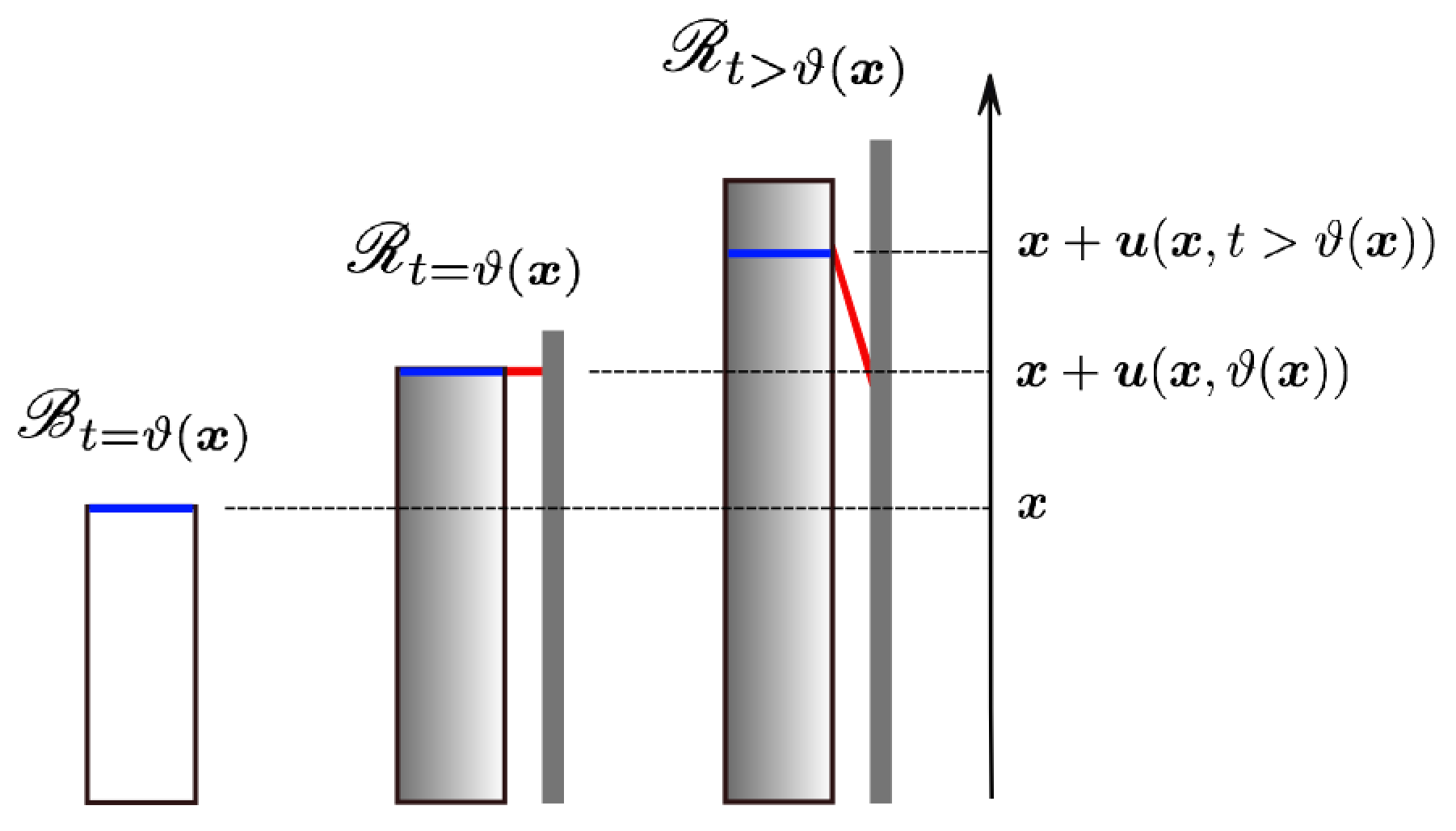}
\caption{\label{15dbars} At the moment of deposition $t=t(x)$, the point $x$ (blue line) in the reference state occupies position $x+\mathring{u}(x))$ in the actual state. A shear spring (red line) is created at this instant, linking point $x$ to the background. At time  $t>\vartheta(x)$ the point $x$ is no longer at the accretion surface and the spring is no longer undeformed.}
\end{figure}

\emph{Surface growth.} Without loss of generality we can assume that the connecting springs are in a stress-free state at deposition so that  $ \mathring{u}(x))=u(x,\vartheta(x)) $ and $q(x,\vartheta(x))=0$, see Fig. \ref{15dbars}. Suppose also that  the end $x=0$ is constrained and that  the bar is exposed to the foundation-unrelated  body forces $b(x)$ and the tractions $s(x)$ on the growing end.  Then the  equilibrium equations at time $t$  read
\beq\label{equiplast15D}
\left\{
\begin{array}{llll}
\partial_x\s(x,t) + k(\mathring{u}(x) - u(x,t)) + b(x) = 0 & \Body_t\\
\s(x,t) = E(\partial_x u(x,t) - \e_p(x)) & \Body_t\\
\s(\psi(t),t) = s(\psi(t)) & \Omega_t\\
u(0,t) = 0 & \Omega_d. 
\end{array}
\right.
\eeq
If $\e_p(x)$ is prescribed, this linear problem  can be solved analytically, in particular we obtain  explicit representations
\beq\label{u015d}
\begin{array}{llll}
\displaystyle\mathring{u}'(x)&=\e_p(x) + \displaystyle\frac{\beta s(x) + (s'(x)+ b(x))\tanh(\beta x)}{\beta E}, \\
\displaystyle\eta (x) &= \displaystyle - \frac{\beta s(x) + (s'(x)+ b(x))\tanh(\beta x)}{\beta E}. 
\end{array}
\eeq
If, instead, $\e_p(x)$ is unknown, we need to prescribe a supplementary condition on the growing end.  For instance, by prescribing $\mathring{u}$,  as we did  in the 1D case,  we obtain 
\beq\label{ep15d}
\e_p=\mathring{u}' - \frac{s}{E} - \frac{\tanh(\beta x)}{\beta E} (s'+ b).
\eeq

\emph{Incremental approach.} It is instructive to see how the incremental approach works in the 1.5D case. We can again define  
\beq\label{incrementalus}
\begin{array}{lll}
u(x,t) = \mathring{u}(x) + \int_{\vartheta(x)}^t\dot u(x,z)\,dz\\
\s(x,t) = s(x) + \int_{\vartheta(x)}^t\dot\s(x,z)\,dz. 
\end{array}
\eeq
and formulate the incremental equilibrium problem   
\beq\label{eq15ddot}
\left\{
\begin{array}{lccc}
\partial_x\dot\s(x,t) -  k \dot u(x,t) = 0   & \Body_t\\
\dot\s(x,t)=E\,\partial\dot u(x,t)   & \Body_t\\
\dot u(0,t)=0  & \Omega_d. 
\end{array}
\right.
\eeq
The boundary condition on the growing end takes the form 
\beq\label{Trincher1D}
\dot\s(x,t)\big{|}_{t=\vartheta(x)}=D_0(x)\left(s'(x)+b(x)\right). 
\eeq
The incremental problem \eqref{eq15ddot}-\eqref{Trincher1D} is the 1.5D counterpart of the 3D problem \eqref{dotred}-\eqref{Trincher}. If we set  $\psi\equiv t$ for simplicity, so that $\vartheta(x)=x$ and $D_0=1$, we can write the incremental solution explicitly
\beq\label{dotuu}
\dot u(x,\psi) = \text{sech}(\beta\psi)\sinh(\beta x) (s'(\psi)+b(\psi))/(E \beta). 
\eeq
After computing the incremental stress from \eqref{eq15ddot}$_2$, we obtain from \eqref{incrementalus}$_2$ the explicit representation of the stress  field
\beq\label{stresshard}
\s(x,\psi) = s(x) + \text{cosh}(\beta x)\int_x^{\psi}(s'(z)+b(z))\,\text{sech}\,(\beta z)\,dz. 
\eeq
Note  that we found the stress  without knowing either the inelastic strain $\e_p(x)$ or the displacement $\mathring{u}(x)$.  

 If $\e_p$ is known, we can now integrate $\s(x,\psi)=E(\partial u(x,\psi) - \e_p(x))$  with a boundary condition $u(0,\psi)=0$ and find  the displacement field $u(x,\psi)$. This gives us  $\mathring{u}(\psi)=u(\psi,\psi)$ in full agreement with  \eqref{u015d}.  If $\e_p(x)$ is unknown  but either $\mathring{u}(x)$ or $D(x)$ are  prescribed, the same Eq. \eqref{u015d}   can be used to compute the inelastic strain.

\section{Illustrations}

\emph{1.5D printing.} Consider a printing device that can control the ratio $D/D_0$ during deposition and assume  for simplicity the bar is elastic, so that  $\e_p=0$. In this case the inelastic effects are due exclusively to the external pre-strain $\mathring{u}'$.

\begin{figure}[!th]
\includegraphics[scale=.3]{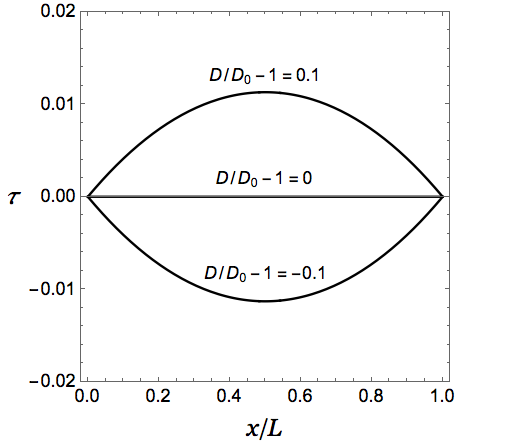}
\caption{\label{DDtau} Residual stresses in a 1.5D bar printed with controlled mismatch between Lagrangian and Eulerian velocities of the growing surface. Here $\beta=1$ and $L=1$.}
\end{figure}

If  the target is a prescribed distribution of residual stress $\bar\tau(x)$   satisfying $\bar\tau(0)=\bar\tau(L)=0$, we can use  \eqref{15Dsecond} to compute the target incompatibility 
\beq
\bar\eta = \left(\bar\tau'' - \beta^2\bar\tau\right)/k. 
\eeq
The corresponding deposition protocol is then given by the equation
\beq\label{Dtau}
D/D_0 - 1 =  \left(\beta^2\bar\tau - \bar\tau''\right)/k. 
\eeq
The inverse problem can be solved similarly. Thus, if we assume that $D/D_0$ is given and integrate \eqref{Dtau} with boundary conditions $\tau(0)=\tau(L)=0$, we obtain the resulting distribution of residual stresses. If, for example, the velocities $D,D_0$ are both constant, we  obtain
\beq
\tau=\left(\small{\frac{D}{D_0}}-1\right)(1-\cosh(\beta x) + \frac{\sinh(\beta x)\tanh(\beta L/2))}{\beta^2}. 
\eeq
This stress distribution, illustrated  in Fig.\ref{DDtau},  vanishes if $D=D_0$, however, when $D>D_0$   the grown bar is left in a state of residual traction while  for $D<D_0$, the residual stresses are  compressive.

\emph{Brick tower.} Consider a  tower   built by continuous deposition of ``bricks'' on one of its ends, while the other end is  fixed. The tower is supported by  a vertical  wall to which the upcoming bricks are attached. We first illustrate the role of the body force distribution $b(x)$. 

We compare two protocols: when the forces $b$ are present in the process of growth, and when they are introduced only after the growth process is completed. We can think about one brick tower manufactured vertically (under the continuous action of gravity) and another brick tower manufactured horizontally (in absence of gravity), and then turned vertically at the end of manufacturing.  In both cases we assume that $s=0$ on the growing end, and neglect plastic strains in the bar.

In the first case (vertical tower) we  have  $b=-\rho g$, where $\rho$ is the linear mass density and $g$ is the acceleration of gravity. Then, from \eqref{inc15} and \eqref{u015d} we obtain
\beq
\eta(x) = \frac{\rho g \tanh(\beta x)}{\beta E}, 
\eeq
and, from \eqref{stresshard}, the final stress distribution reads
\beq\label{sigma1}
\s(x) = \frac{\rho g}{\beta}\text{cosh}(\beta x)\left(\text{gd}(\beta x) - \text{gd}(\beta L)\right),
\eeq
where $\text{gd}(x)=2\,\text{arctan}(\text{tanh}(x/2))$. In the limit  $\beta\rightarrow 0$, when the foundation can be neglected,  we obtain an elastic growth with $\eta(x)=0$ and $\s(x)=-\rho g (L-x)$. In the limit  $\beta\rightarrow \infty$ the foundation dominates and carries the applied loads, so that the bar is stress free. 
\begin{figure}[!th]
\includegraphics[scale=.36]{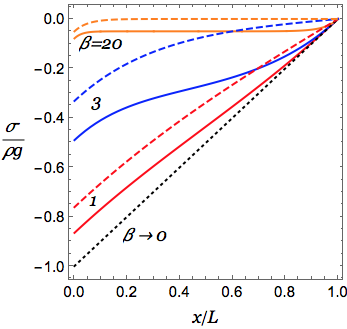}\,\,
\includegraphics[scale=.15]{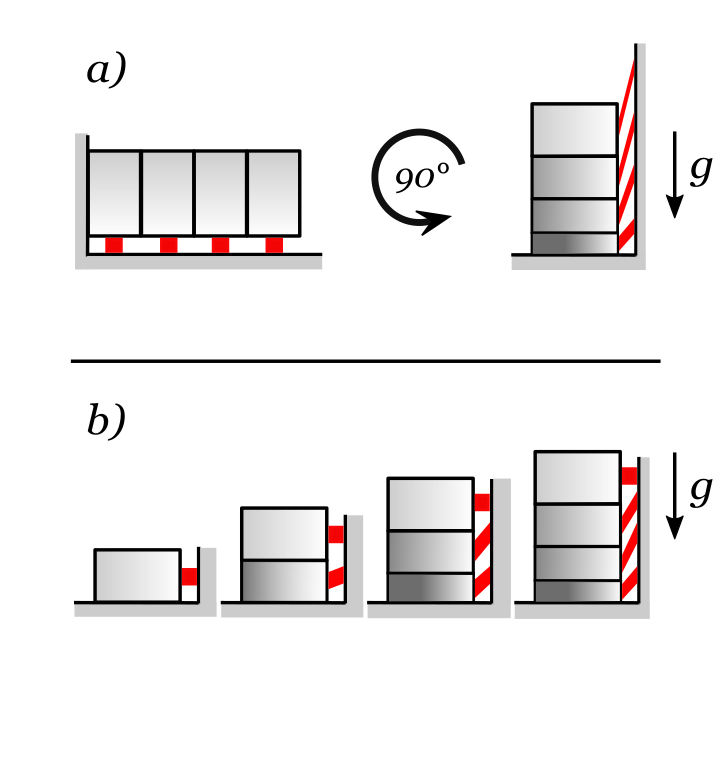}
\caption{\label{walls} Dashed curves: stress distribution in the structure  subjected to   gravity after it is  manufactured, see also (a). Continuous curves: stress in the structure  manufactured under the action of gravity, see also ( b). The curves are distinguished  by the value of the parameter $\beta$. The dotted line represents the stress in the limiting case when there is no foundation.}
\end{figure}

In the second case, we manufacture the tower without gravity while assuming that  $D=D_0$, so that  $\eta=0$. 
If we now rotate the structure  (switch on gravity)  the stress  distribution can be found from  \eqref{15Dfirst} with boundary conditions $\s(L)=0$ and $u(0)=0$. With $b=-\rho g$ we obtain 
\beq\label{sigma2}
\s(x)=-\frac{\rho g}{\beta}\text{sech}(\beta L)\text{sinh}(\beta(L-x)). 
\eeq
Also in this case, the purely elastic solution is  recovered in the   statically determined   limit $\beta\rightarrow 0$, while  in the limit $\beta\rightarrow \infty$ the stress in the bar drops to zero, as the applied load is  fully  carried by the foundation. 

The stress distributions  in the two brick towers,   one manufactured vertically and the other one manufactured horizontally and then rotated, are illustrated  in Fig.\ref{walls}. A comparison  of  the stress profiles at  the same values of $\beta$ shows that the   distributions are different, which highlights the inherent path dependence of the process of incompatible surface growth. 
 \begin{figure}[!th]
\includegraphics[scale=.29]{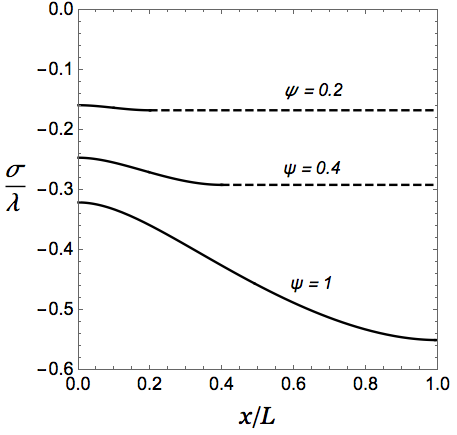}
\includegraphics[scale=.18]{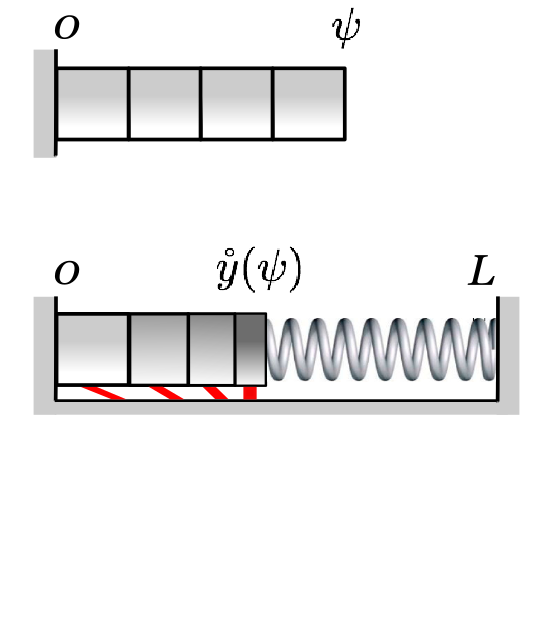}
\caption{\label{freezing} {\it Left:}  stress in the bar (solid) and in the spring (dashed) for different positions $\psi/L$ of the interface, for $\beta=3$ and $\alpha=1$. {\it Right:}  Reference (up) and current (down) configurations of the the growing bar, for a specific value of $\psi$.}
\end{figure}

\bigskip
  
\emph{Growth against an elastic constraint.} Finally,  consider a bar growing against an elastic constraint while the non-growing end is kept fixed. Also in this case we assume for analytical transparency that $\e_p=0$ and neglect body forces. The traction $s$ exerted by the elastic constraint depends on the current position  $\mathring{y}(\psi) = \psi+\mathring{u}(\psi)$ of the accreting surface, so if $s=\bar s(\psi+\mathring{u}(\psi))$ is known, \eqref{u015d}$_1$ can be viewed as a first order ordinary differential equation for $\mathring{u}$, that can be integrated with an initial condition $\mathring{u}(0)=0$. For instance, if the growth resisting elastic force is linear  $ \bar s(z) = - \lambda z/L, $
where $\lambda$ is an elastic constant, integration gives 
\beq
\begin{array}{lll}
&\mathring{u}(x)=L\alpha^{-1} - x \\
&- e^{xLm\beta^2}L\alpha^{-1}\beta^m(\beta L \cosh(\beta x) + \alpha\sinh(\beta x))^m
\end{array}
\eeq
where we have set  $\alpha=\lambda/E$ and $m=\alpha^2/(\alpha^2-L^2\beta^2)$. We can now  compute the surface traction $s $ and,  by making use of \eqref{stresshard},  obtain the stress in the system. The results, illustrated in Fig.\ref{freezing}, show that surface growth in this setting can produce  highly inhomogeneous stress fields.

\section{Conclusions}

In this paper we have developed a  non-incremental theory of inelastic surface growth allowing one to compute  the distribution of plastic strains that were underplayed   in the incremental approach presented in \cite{ZT17}. To illustrate the new ideas, we  presented several analytically transparent  examples where the general theory was applied  to the growth of bars  attached to a  Winkler foundation. By explicit solutions we demonstrated  the path-dependence of the process of inelastic growth  and highlighted  different roles played in the theory by compatible and incompatible inelastic strains. Surface growth in such 1.5D systems is of interest in various applications, for instance, in civil engineering of  high-rise construction, in the study of self-propulsion of cells on rigid substrates and in  micro-technologies  based on ``de-peeling''  of programmable structures with non-trivial relaxed shapes.

\end{document}